\documentclass[12pt]{article}
\usepackage{cite}
\usepackage{epsf}
\usepackage{psfig}
\usepackage{times}

\makeatletter

\def\paragraph{\@startsection{paragraph}{4}{\z@}{+2.00ex plus
 +1ex minus +.2ex}{1.5ex plus .2ex}{\it\normalsize}}

\def\section{\@startsection {section}{1}{\z@}{+3.0ex plus +1ex minus
  +.2ex}{2.3ex plus .2ex}{\normalsize\bf\boldmath}}
\def\subsection{\@startsection{subsection}{2}{\z@}{+2.5ex plus +1ex
minus +.2ex}{1.5ex plus .2ex}{\normalsize\bf\boldmath}}
\def\subsubsection{\@startsection{subsubsection}{3}{\z@}{+3.25ex plus
 +1ex minus +.2ex}{1.5ex plus .2ex}{\normalsize\bf\boldmath}}

\expandafter\ifx\csname mathrm\endcsname\relax\def\mathrm#1{{\rm #1}}\fi

\newcounter{saveeqn}

\@addtoreset{equation}{section}

\newcommand{\denone}{y_{\rm min}-\delta y-1+y_{\rm X}}
\newcommand{\dentwo}{y_{\rm X}-2-y_{\rm min}\*y_{\rm X}+ 
      y_{\rm X}^2-2\*y_{\rm min}+y_{\rm min}^2+2\delta y+
      y_{\rm X}\delta y+\delta y^2-2y_{\rm min}\delta y}

\newcommand{\rmt}{\rm\textstyle}
\newcommand{\rms}{\rm\scriptstyle}
\newcommand{\nub}{\overline{\nu}}
\newcommand{\nubar}[0]{\overline{\nu}}

\newcommand{\txnunub}[1]{\nu_{#1},\nub_{#1}}

\newcommand{\stw}{\mbox{$\sin^2\theta_W$}}
\newcommand{\stwos}{\mbox{$\sin^2\theta_W^{\rms(on-shell)}$}}

\newcommand{\Rnu}{\mbox{$R^{\nu}$}}
\newcommand{\Rnub}{\mbox{$R^{\nub}$}}

\newcommand{\Rmeas}{\mbox{$R_{\rms exp}$}}
\newcommand{\Rmeasnu}{\mbox{$R_{\rms exp}^{\nu}$}}
\newcommand{\Rmeasnub}{\mbox{$R_{\rms exp}^{\nub}$}}

\newcommand{\qbar}[0]{\overline{q}}

\newcommand{\sbar}[0]{\overline{s}}

\newcommand{\Sbar}[0]{\overline{S}}

\newcommand{\uubar}[0]{\mbox{$\stackrel{(-)}{u}$}}
\newcommand{\ddbar}[0]{\mbox{$\stackrel{(-)}{d}$}}

\newcommand{\savbar}[0]{\langle\overline{s}(x)\rangle}

\newcommand{\sav}[0]{\langle{s(x)}\rangle}

\begin{document}

\pagestyle{empty}

\vspace*{1cm}
\begin{center}

{\Large \bf  
Conventional Physics Explanations for the NuTeV~\stw
 }

\vspace*{1cm}

{\sc Kevin~S.~McFarland$^{\star}$ and Sven-Olaf Moch$^{\dagger}$}

\vspace*{.5cm}

{\normalsize \it
$^{\dagger}$~Deutsches Elektronensynchrotron DESY \\
Platanenallee 6, D--15738 Zeuthen, Germany \\[1ex]
$^{\star}$~University of Rochester, Rochester, NY 14627 USA}
\par
\end{center}
\vskip 1cm
\begin{center}
\bf Abstract
\end{center} 
{\it The NuTeV experiment has measured 
$\stw^{({\rms on-shell)}}=0.2277\pm0.0013({\rmt stat})\pm0.0009({\rmt syst})$,
approximately 3 standard deviations above the standard model prediction.  
This discrepancy has motivated speculation that the NuTeV
result may be affected significantly by neglected experimental or
theoretical effects.  We examine the case for a number of proposed
explanations.
}
\par
\vskip 1cm

\section{Introduction}

The observable $R^{-}$
 \begin{eqnarray}
R^{-} &\equiv& \frac{\sigma(\nu_{\mu}N\rightarrow\nu_{\mu}X)-
                   \sigma(\nub_{\mu}N\rightarrow\nub_{\mu}X)}
                  {\sigma(\nu_{\mu}N\rightarrow\mu^-X)-  
                   \sigma(\nub_{\mu}N\rightarrow\mu^+X)} , 
\label{eqn:rminus}
\end{eqnarray}
was first suggested by Paschos and Wolfenstein~\cite{Paschos:1973kj}
to measure $\sin^2 \theta_W$.
\begin{eqnarray}
R^{-} &\equiv& {1 \over 2} - \sin^2 \theta_W 
\label{eqn:rminusLO}
\end{eqnarray}
under the assumptions of equal momentum carried by the $u$ and $d$
valence quarks in the target and of equal momentum carried by the
heavy quark and anti-quark seas.  The Paschos-Wolfenstein numerator
and denominator are independent of the sea quark momenta since
$\sigma^{\nu q}=\sigma^{\nub\, \qbar}$ and $\sigma^{\nub
q}=\sigma^{\nu \qbar}$.  $R^-$ is more difficult to measure than 
the ratio of neutral current to charged current cross-sections, 
$R^\nu$, primarily because the neutral current scattering of $\nu$
and $\nub$ yield identical observed final states which can only be
distinguished through {\em a priori} knowledge of the initial state
neutrino.  Therefore, the measurement of $R^-$ requires separated
neutrino and anti-neutrino beams.

As a test of the electroweak predictions for neutrino nucleon
scattering, NuTeV has performed a single-parameter fit using its $R^-$
data to $\stw$ with all other parameters assumed to have their
standard values, e.g., standard electroweak radiative corrections with
$\rho_0=1$.  This fit determines
\begin{eqnarray}
    \sin^2\theta_W^{({\rms on-shell)}}&=&0.22773
                \pm0.00135({\rmt stat.})\pm0.00093({\rmt syst.})
        \nonumber\\
        &-&0.00022\times(\frac{M_{\rms top}^2-(175 \: \mathrm{GeV})^2}
                               {(50 \: \mathrm{GeV})^2})
        \nonumber\\
        &+&0.00032\times \ln(\frac{M_{\rms Higgs}}{150 \:
\mathrm{GeV}}).
\label{eqn:result}
\end{eqnarray}
The small dependences in $M_{\rms top}$ and $M_{\rms Higgs}$ result
from radiative corrections as determined from code supplied by
Bardin\cite{Bardin:1986bc} and from V6.34 of
ZFITTER\cite{Bardin:1999yd}; however, it should be noted that these
effects are small given existing constraints on the top and Higgs
masses\cite{Abbaneo:2001ix}.  A fit to the precision electroweak data,
excluding neutrino measurements, predicts a value of
$0.2227\pm0.00037$\cite{Abbaneo:2001ix},
approximately $3\sigma$ from the NuTeV measurement.

The experimental details and theoretical treatment of cross-sections
in the NuTeV electroweak measurement are described in detail
elsewhere~\cite{Zeller:2001hh}. In brief, NuTeV measures the
experimental ratio of neutral current to charged current candidates in
both a neutrino and anti-neutrino beam.  A Monte Carlo simulation is
used to express these experimental ratios in terms of fundamental
electroweak parameters.  This procedure implicitly corrects for
details of the neutrino cross-sections and experimental backgrounds.
For the measurement of $\stw$, the sensitivity arises in the $\nu$
beam, and the measurement in the $\nubar$ beam is the control sample
for systematic uncertainties, as suggested in the Paschos-Wolfenstein
$R^-$ of Eqn.~(\ref{eqn:rminus}).  For simultaneous fits to two
electroweak parameters, e.g., $\stw$ and $\rho$ or left and right
handed couplings, this control of systematics by the
Paschos-Wolfenstein $R^-$ cannot be realized.

\subsection{Differences between NuTeV and $R^{-}$}
\label{sect:notR-}

NuTeV does not measure exactly $R^-$ but rather
combinations of experimentally observed neutral-current to
charged-current cross-section ratios, $R^{\rm meas}$ in neutrino and
anti-neutrino beams.  Note that $R^-$ itself can actually be written as
\begin{eqnarray}
R^{-} &\equiv& \frac{\sigma(\nu_{\mu}N\rightarrow\nu_{\mu}X)-
                   \sigma(\nub_{\mu}N\rightarrow\nub_{\mu}X)}
                  {\sigma(\nu_{\mu}N\rightarrow\mu^-X)-  
                   \sigma(\nub_{\mu}N\rightarrow\mu^+X)} , \\
 &=& \frac{\frac{\sigma(\nu_{\mu}N\rightarrow\nu_{\mu}X)}{\sigma(\nu_{\mu}N\rightarrow\mu^-X)}-
           \frac{\sigma(\nub_{\mu}N\rightarrow\nub_{\mu}X)}{\sigma(\nub_{\mu}N\rightarrow\mu^+X)}  
           \frac{\sigma(\nub_{\mu}N\rightarrow\mu^+X)}{\sigma(\nu_{\mu}N\rightarrow\mu^-X)}}  
                  {1-  
              \frac{\sigma(\nub_{\mu}N\rightarrow\mu^+X)}{\sigma(\nu_{\mu}N\rightarrow\mu^-X)}} , \\
 &\equiv& \frac{R^\nu - r R^{\nub}}{1-r},
\label{eqn:rminusRs}
\end{eqnarray}
where $R$ is the ratio of neutral-current to charged-current total
cross-sections and $r$ is the ratio of charged-current anti-neutrino
to neutrino total cross-sections.  Therefore, $R^-$ itself is a
combination of $R$ from neutrino and anti-neutrino beams.

NuTeV does not measure the total cross-section ratios of
neutral and charged current interactions.  The ratios actually
measured by NuTeV, 
\begin{equation}
\Rmeasnu= 0.3916 \pm 0.0007~{\rm and}~\Rmeasnub= 0.4050 \pm 0.0016,
\end{equation}
include non-muon neutrino backgrounds, the effects of experimental
cuts, cross-talk between candidates in the numerator and denominator
and final state effects.

The electron neutrino background can be reliably subtracted as
discussed below in Section~\ref{sect:expConcerns}, and the effects of
experimental cuts and cross-talk, although large, can be controlled
experimentally and introduce no large corrections to the $R^-$
dependence on $\stw$ as will be shown in Section~\ref{sect:expModel}.

There are two final state effects which make modest corrections to the
NuTeV observables.  The first are small differences in acceptance in
neutral and charged-current events due to the effect of the final
state muon.  The second is the effect of the final state in semi-leptonic charm
decay.  In such decays, some of the final state charmed hadron's
energy is carried away in final state neutrinos and some may appear as
final state muons.  Further complicating the decay, those muons may
affect whether the event is measured as a charged-current candidate or
neutral-current candidate.  These effects are small in the final
analysis, but they do make some correction to the NuTeV observables
not present in $R^-$.  Both of these effects are fully modeled and
corrected for in the NuTeV analysis.

A more significant difference between NuTeV's analysis and $R^-$ is
the way that $\stw$ is extracted from the two $\Rmeas$.  The analysis
uses a fit to two parameters, $\stw$ and the effective charm mass, $m_c$, in
charged-current charm production, which is the largest theoretical
uncertainty in the analysis.  This $m_c$ is externally constrained
from the NuTeV measurements of charged-current charm production in the
two muon final state, and so a 1C fit determines the result of
Eqn.~(\ref{eqn:result}).  NuTeV has also performed a fit without this
external charm mass constraint~\cite{Zeller:2002he}, and has found
\begin{eqnarray}
    \sin^2\theta_W^{({\rms on-shell)}}&=&0.22738
                \pm0.00164({\rmt stat.})\pm0.00076({\rmt syst.}),
\label{eqn:result0C}
\end{eqnarray}
which is consistent with the 1C fit result.

\begin{figure}[tbp]
\begin{center}
\epsfxsize=0.8\textwidth\epsfbox{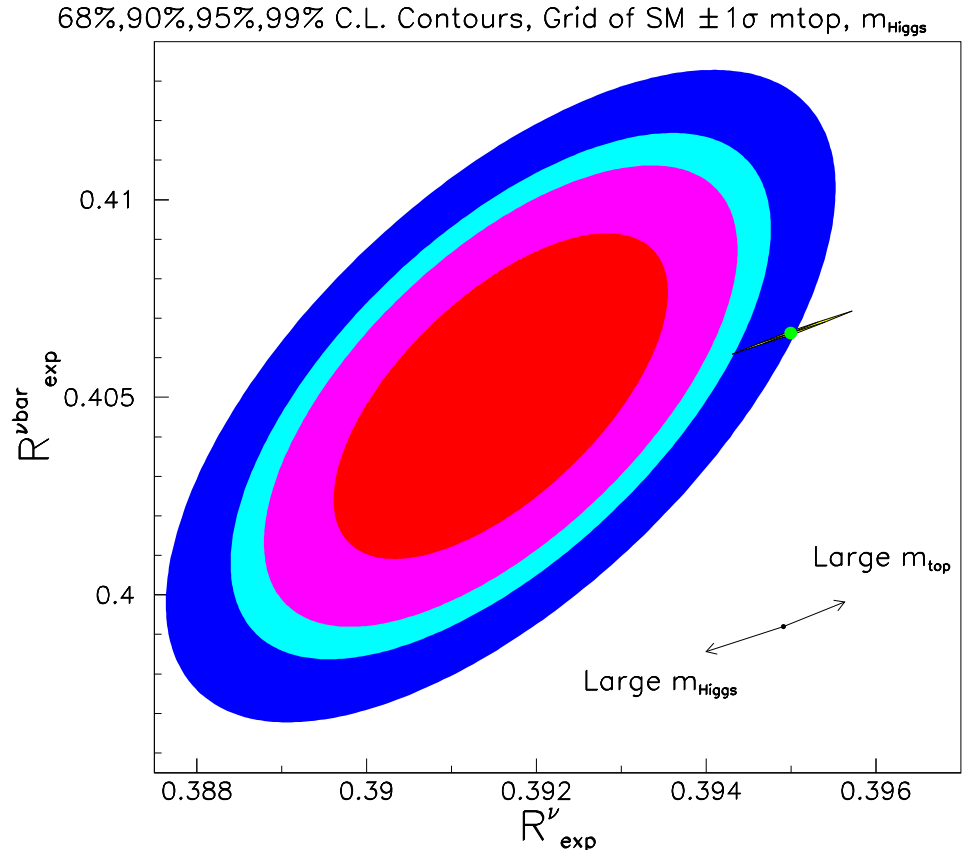}
\end{center}
\caption{\it
  The measurements of $\Rmeasnu$ and $\Rmeasnub$, shown as an error
  ellipse.  The uncertainties in the error ellipse include theoretical
  uncertainties in relating $\Rmeasnu$ and $\Rmeasnub$ to fundamental
  electroweak parameters, which result in the correlation between the
  two measurements.  Note that the $\Rmeasnub$ is in agreement with
  the Standard Model expectation, shown as the point, whereas the
  $\Rmeasnu$ measurement is not.}
\label{fig:Rnunub}
\end{figure}

It is instructive to decompose these (linear) fits into combinations
of $\Rmeasnu$ and $\Rmeasnub$ in the spirit of
Eqn.~(\ref{eqn:rminusRs}).
Writing
\begin{equation}
    \tilde{R}^- \equiv \frac{\Rmeasnu - a \Rmeasnub}{b},~~~
     \ni^\prime \frac{d \tilde{R}^-}{d\stw} = -1
\label{eqn:pseudoR}
\end{equation}
where the latter equality holds by choice of $b$,
we find that for the 1C fit $a=0.2492$, $b=0.6170$, and for the 0C fit 
$a=0.4526$, $b=0.6116$.   The similarity between the $b$ values for
the two fits is a statement that the sensitivity to $\stw$ almost
entirely resides in $\Rmeasnu$ and not $\Rmeasnub$, and the near
equivalence of the 1C and 0C results indicates that $\Rmeasnub$ is in
agreement with expectations as is illustrated in
Figure~\ref{fig:Rnunub}.  The discrepancy with the Standard Model of
these results comes then from the fact that $\Rmeasnu$ is not as expected.
Note that these separate measurements, either cast as $\Rmeasnu$ and
$\Rmeasnub$ or as the 1C and 0C fits, will constrain any attempt to
explain the NuTeV $\stw$ that seeks to make large changes in $\Rnu$
and $\Rnub$ separately to effect a large change in $R^-$.

\section{Experimental Issues}
\label{sect:expConcerns}

One of the largest experimental corrections to the NuTeV analysis is
the subtraction of electron neutrino charged-current events from the
neutral current candidate sample.  Approximately $5\%$ of the neutral
current candidates are electron neutrino charged
currents~\cite{Zeller:2001hh}, and so a $20\%$ overestimate of this
rate would be sufficient to explain the difference between the NuTeV
$\stw$ and the Standard Model expectation.

The dominant source of electron neutrinos in the NuTeV beams are
$K^\pm_{e3}$ decays.  NuTeV determines the electron neutrino
background by two methods: one an indirect determination using a beam
Monte Carlo tuned to the neutrinos observed from $K^\pm_{\mu 2}$
decays, and the other a direct, but less precise, measurement of the
electron neutrinos~\cite{Avvakumov:2002jj}.  An important check is
that the direct and Monte Carlo methods agree, and in fact the ratio
of measured to Monte Carlo predicted events is $1.05\pm0.03$ in the
neutrino beam and $1.01\pm0.04$ in the anti-neutrino beam.

The largest uncertainty in the Monte Carlo method is the $1.4\%$
fractional uncertainty in the $K^\pm_{e3}$ branching
ratio~\cite{Hagiwara:2002fs}.  An interesting recent development comes
from the BNL-E865 experiment which has recently measured a branching
ratio for $K^\pm_{e3}$~\cite{Sher:2003fb} that is $6\%$ larger than
the value used by NuTeV.  If this result is correct, it is interesting
to note that it would not disrupt the agreement between the direct and
Monte Carlo measurements of the $\nu_e$ rate at NuTeV and that it
would in fact {\em increase} the discrepancy of the NuTeV $\stw$ with
the prediction by slightly less than one standard
deviation~\cite{Zeller:2002he}.

\begin{figure}[tbp]
\epsfxsize=0.8\textwidth\epsfbox[20 150 530 690]{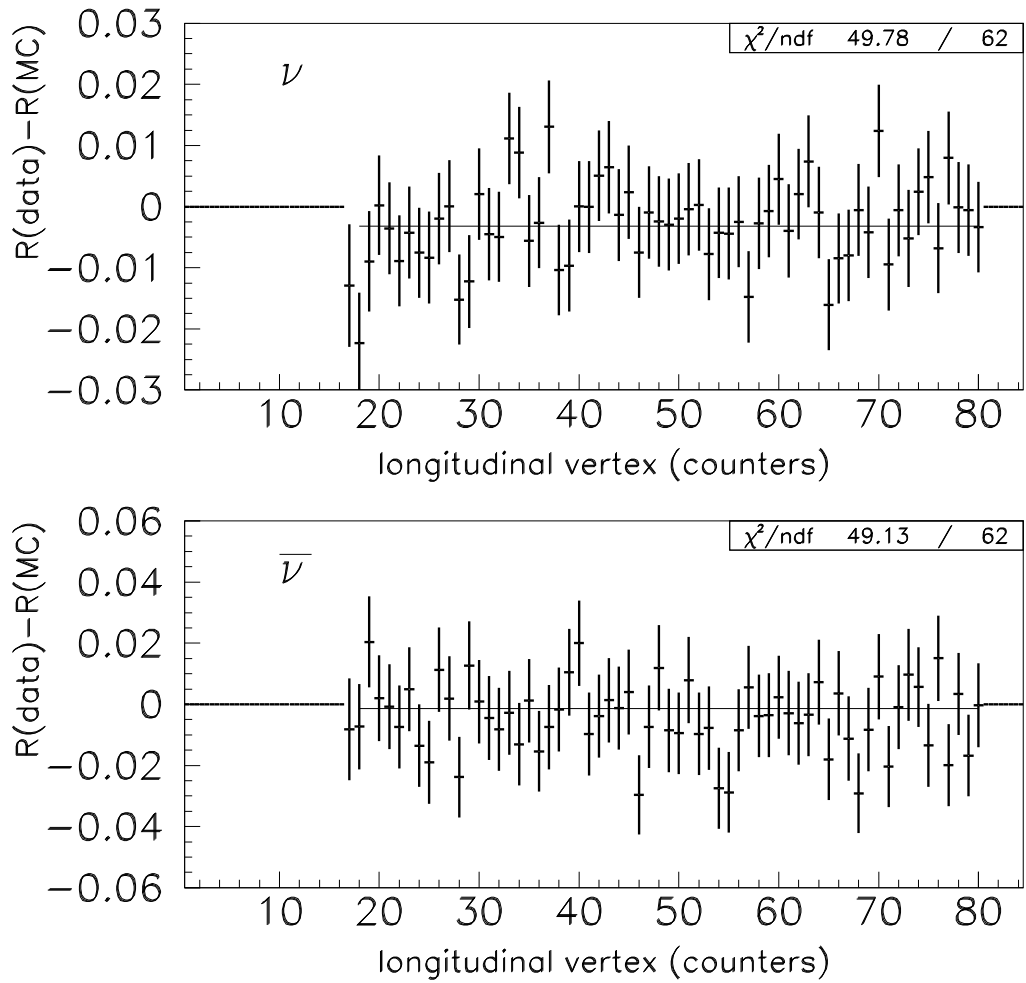}
\caption{\it
  The $\Rmeas$ in the neutrino and anti-neutrino beams as a function
of the depth of the neutrino interaction within the detector along the
beam direction.}
\label{fig:place}
\end{figure}

\begin{figure}[tbp]
\begin{center}
\epsfxsize=0.75\textwidth\epsfbox[20 150 530 620]{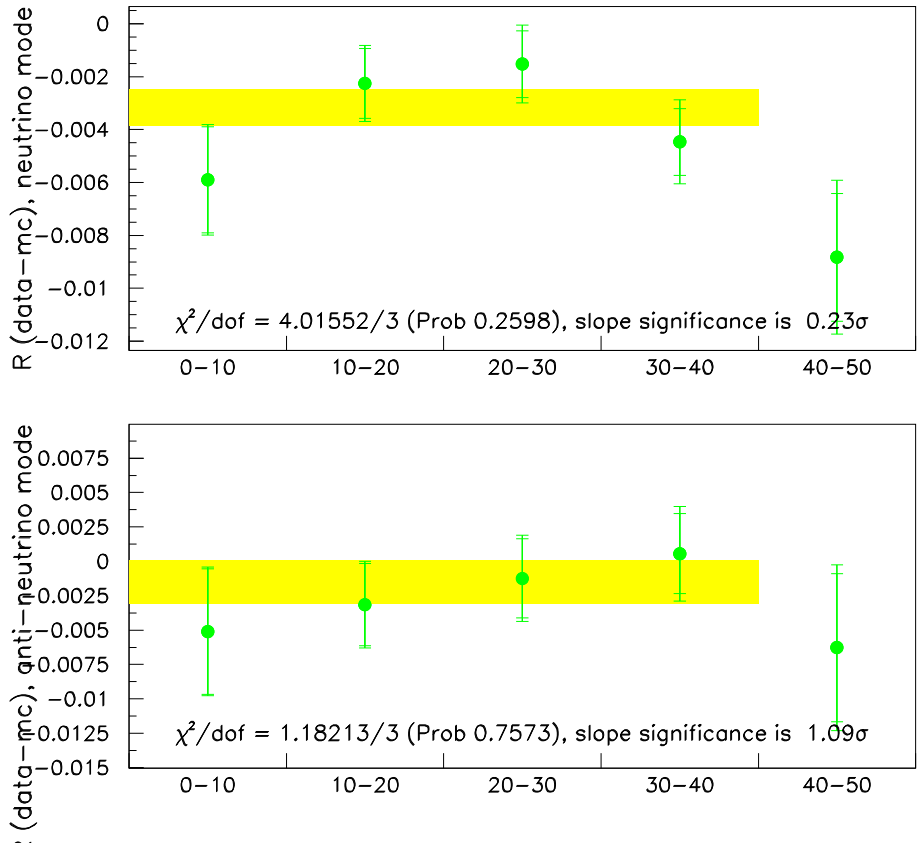}
\end{center}
\caption{\it
  The NuTeV $\Rmeas$ binned in square annuli of transverse position
from the center to the outer part of the detector.  The first four
bins are used in this analysis.}
\label{fig:vertex}
\end{figure}

\begin{figure}[tbp]
\mbox{
\epsfxsize=0.45\textwidth\epsfbox{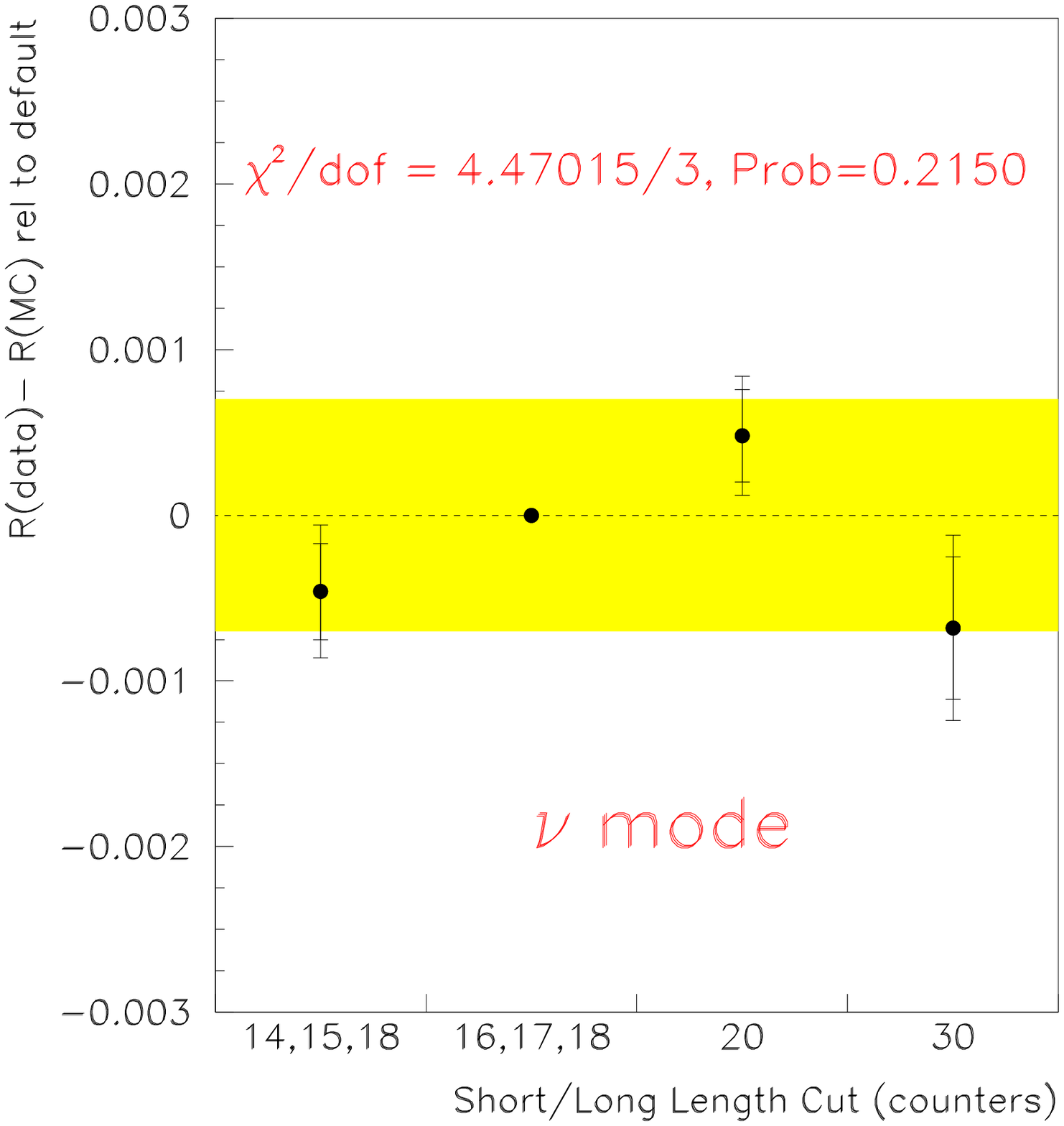}
\hspace*{0.2in}
\epsfxsize=0.45\textwidth\epsfbox{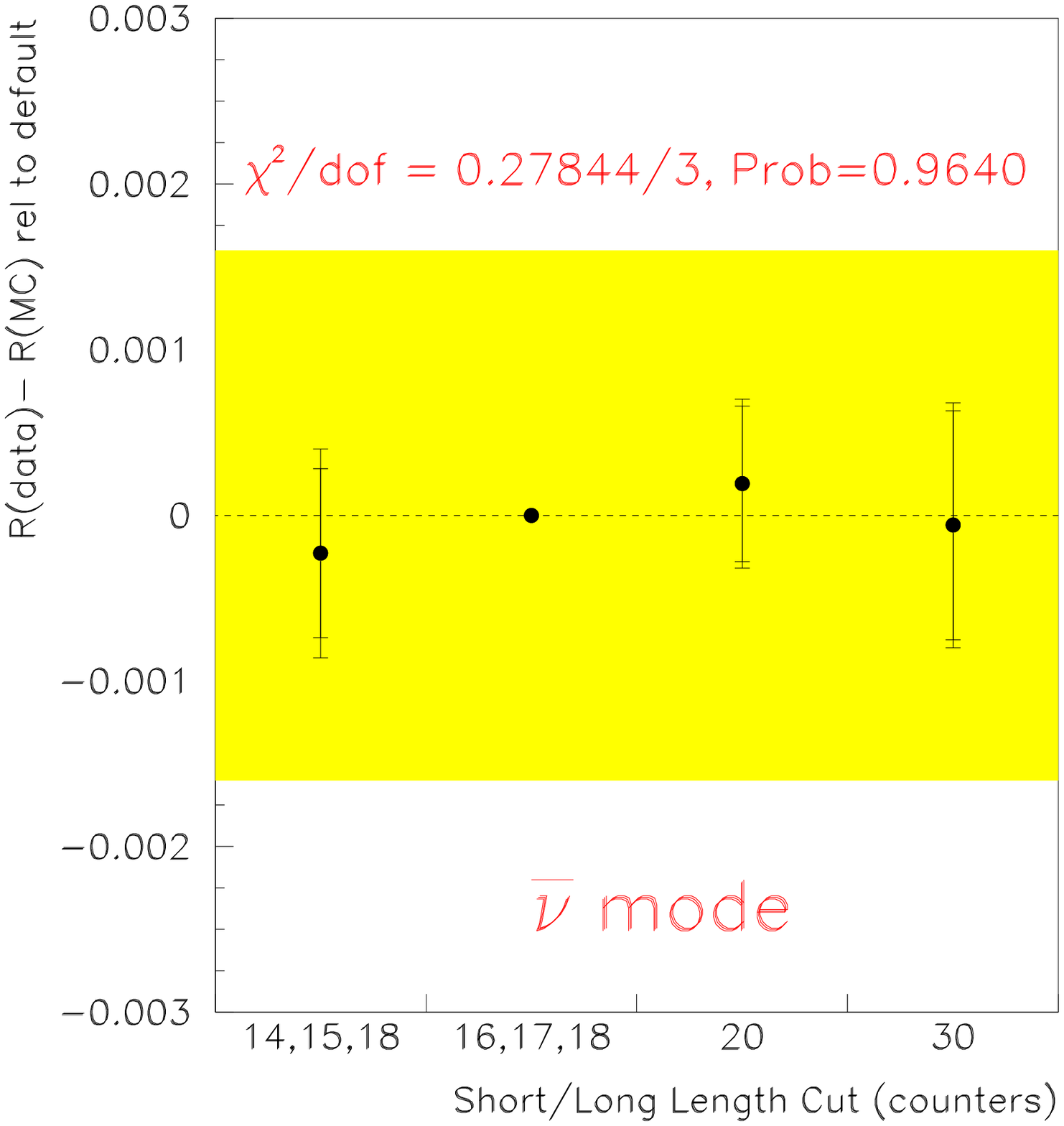}
}
\caption{\it
  The effect of varying the NuTeV neutral-current/charged-current
  separation cut on $\Rmeas$ in the neutrino (left) and anti-neutrino (right)
  beams.  The error bars represent the statistical uncertainty on the
  {\em charge}.  The yellow band is the statistical uncertainty of the
  ratio as measured from the whole sample.}
\label{fig:lcut}
\end{figure}

\begin{figure}[tbp]
\epsfxsize=0.90\textwidth\epsfbox[15 145 522 660]{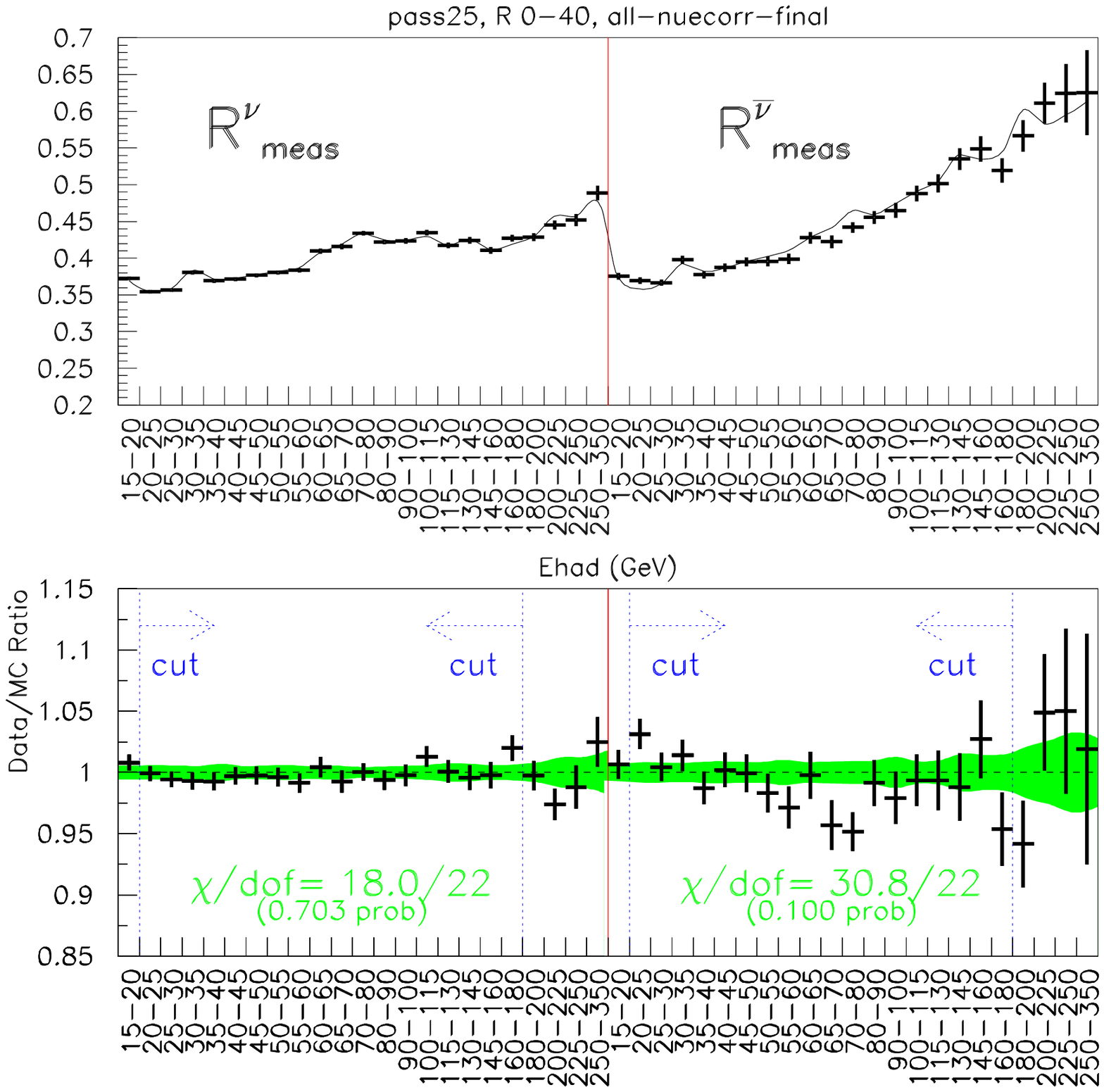}
\caption{\it
  The $\Rmeas$ as a function of visible energy in the neutrino
  and anti-neutrino beams.  The ratio of data to Monte Carlo is shown
below, with a green band to represent the systematic uncertainties in
the comparison.}
\label{fig:rvsE}
\end{figure}

The NuTeV observables, $\Rmeasnu$ and $\Rmeasnub$, have been studied
as a function of position within the detector (Figures~\ref{fig:place}
and \ref{fig:vertex}), the containment length used to separate neutral
and charged current candidates (Figure~\ref{fig:lcut}) and the visible
energy in the detector (Figure~\ref{fig:rvsE}).  Within the
statistical and systematic uncertainties of these comparisons, no
unexpected deviations as a function of event variables are observed.

\section{Electroweak Radiative Corrections}

NuTeV's analysis includes complete one-loop electroweak radiative
corrections~\cite{Bardin:1986bc,Bardin:1999yd}.  These corrections can
be separated into those that can be absorbed into effective weak
neutral current $\nu$-$q$ couplings and those that cannot.  The former
cause a $-0.00159$ shift in the NuTeV reported $\stw$. The latter
corrections, dominated by a $W$--$\gamma$ box diagram and by
acceptance effects from bremsstrahlung in the final-state muon in
the charged current shift $\stw$ by an enormous $-0.00795$.

One concern about the NuTeV result is that it relies on this single
calculation of the radiative corrections.  This calculation has been
successfully checked against other partial
calculations~\cite{DeRujula:1979jj,Sirlin:1981yz} in the limits where
both are expected to agree; however, another complete calculation as a
cross-check might be advisable.

\section{QCD corrections}

A sizable part of the quoted theoretical error of $\sin^2 \theta_W$
as determined by NuTeV is due to QCD effects. Therefore, it is
mandatory to investigate all potential sources of QCD uncertainties to
the best possible accuracy. Sources of QCD effects include parton
distributions and their uncertainties, perturbative QCD corrections at
higher orders, isospin breaking effects or nuclear effects.  In the
following, we will discuss these in turn.

\subsection{Perturbative QCD corrections}

The QCD corrections to deep-inelastic scattering at higher orders are
long known. For instance, in a complete next--to--leading order (NLO)
treatment, and assuming massless quarks, the Paschos--Wolfenstein
ratio Eqn.~(\ref{eqn:rminus}) becomes\footnote{In
ref.~\cite{Davidson:2001ji} the correct value \cite{Forte:private} for
$\delta C^1-4 \delta C^3$ is $32/9$, with $\delta C^1=C^1-C^2$ and
$\delta C^3=C^2-C^3$, cf. Eqn.~(\ref{eqn:coefffct}).}
\begin{eqnarray}
R^{-} &=& g_L^2 - g_R^2 + \frac{u^- - d^- + c^- - s^-}{u^- + d^-} 
\left( 3(g_{Lu}^2 - g_{Ru}^2) + (g_{Ld}^2 - g_{Rd}^2) +
  \frac{8}{9} \frac{\alpha_s}{\pi} (g_L^2 - g_R^2) \right)
\nonumber\\
& & 
 + {\cal{O}}\left(\frac{1}{(u^- + d^-)^2}\right)
 + {\cal{O}}\left(\alpha_s^2\right)\, .
\label{eqn:rminusNLO}
\end{eqnarray}
Eqn.~(\ref{eqn:rminusNLO}) uses the one-loop coefficient
functions~\cite{Bardeen:1978yd,Furmanski:1982cw} and  
holds, if no assumption on the parton content of the target is made. 
Here $q^- \equiv q-\bar{q}$ is the second Mellin moment of the 
corresponding 
quark and anti-quark distribution, 
\begin{eqnarray}
  \label{eqn:pdfmom}
  q^- \equiv \int_0^1\, dx\, x\, \left( q(x) - \bar{q}(x) \right)\, .
\end{eqnarray}
For consistency, NLO parton distributions are required, of course, 
and the nucleon is assumed to contain four light flavors. We 
will later comment on the treatment of charm as a massive quark.

The result in Eqn.~(\ref{eqn:rminusNLO}) has been expanded in powers
of the dominant isoscalar combination of parton distributions, $u^- +
d^-$. It shows the well-known fact, that the Paschos--Wolfenstein
relation receives corrections, if the target has an isotriplet
component, $u\not=d$, or sea quark contributions have a $C$-odd
component, $s^-\not=0$ or $c^-\not=0$. In particular, the QCD
corrections only affect these isotriplet $u^- - d^-$ and $C$-odd
terms, $s^-, c^-$.  It is worth noting, that the coefficient at order
$\alpha_s$ in Eqn.~(\ref{eqn:rminusNLO}) is the same in the
$\overline{MS}$-scheme and the DIS-scheme. This is due to the relevant
combination of coefficient functions,
\begin{eqnarray}
\label{eqn:coefffct}
-\frac{1}{4} C_{L,\rm q} + C_{2,\rm q} - C_{3,\rm q}\, 
\end{eqnarray}
being invariant under changes between these two schemes. Of course,
the NLO parton distributions entering in Eqn.~(\ref{eqn:rminusNLO})
differ slightly in the two schemes, most notably in the region of
larger $x$. Also, it is worth pointing out, that in the combination of
coefficient functions in Eqn.~(\ref{eqn:coefffct}) all dependence on
the factorization scale $\mu_{\rm f}$ (i.e. all logarithms
$\ln(Q^2/\mu^2_{\rm f})$) cancels.

Numerically, the correction factor $\frac{u^- - d^- + c^- - s^-}{u^- + d^-}$
is $-0.0232$, evaluated using the NuTeV LO PDFs~\cite{Zeller:2001hh}
at the experimental $Q^2$ which is extracted from fits to CCFR
cross-sections~\cite{Yang:2000ju}.  Inclusion of the higher order
terms in $\frac{1}{u^- + d^-}$ changes this to $-0.0213$.  This
evaluation assumes the strange and anti-strange seas carry equal
momentum and that PDFs are isospin symmetric (see discussions in
Sections~\ref{sect:PDFs} and \ref{sect:iso}).  Therefore, the NLO QCD
correction to $R^-$ is $-0.00035$.

It is possible, to extend Eqn.~(\ref{eqn:rminusNLO}) to
next--to--next--to--leading order (NNLO) in QCD, with the known
two-loop coefficient
functions~\cite{vanNeerven:1991nn,Zijlstra:1992kj,Zijlstra:1992qd,Moch:1999eb}.
Putting the renormalization and the factorization scales $\mu_{\rm f}
= \mu_{\rm r} = Q$, we find
\begin{eqnarray}
R^{-} &=& g_L^2 - g_R^2 + \frac{u^- - d^- + c^- - s^-}{u^- + d^-} 
\biggl( 3(g_{Lu}^2 - g_{Ru}^2) + (g_{Ld}^2 - g_{Rd}^2) 
\nonumber\\
& & 
\left.
+
  \left\{ \frac{8}{9} \frac{\alpha_s}{\pi} + \left[ 
      \frac{5551}{810} - \frac{1}{27} \zeta_2 - \frac{16}{45} \zeta_3
      - \frac{83}{162} n_f \right] \frac{\alpha_s^2}{\pi^2} \right\} (g_L^2 - g_R^2) \right)
\nonumber\\
& & 
 + {\cal{O}}\left(\frac{1}{(u^- + d^-)^2}\right)
 + {\cal{O}}\left(\alpha_s^3\right)\, ,
\label{eqn:rminusNNLO}
\end{eqnarray}
which holds in the $\overline{MS}$-scheme. Here, $n_f$ denotes the
number of light (massless) quark flavors, and $\zeta_2=1.644934068$,
$\zeta_3=1.202056903.$ For arbitrary scales $\mu_{\rm f} \neq Q$,
there will be at most single logarithmic dependence on $Q^2/\mu^2_{\rm
f}$ at order $\alpha_s^2$ in Eqn.~(\ref{eqn:rminusNNLO}). All double
logarithms $\ln^2(Q^2/\mu^2_{\rm f})$ cancel due to
Eqn.~(\ref{eqn:coefffct}).  In the DIS-scheme, the coefficient at
order $\alpha_s^2$ changes slightly. There, the expression in square
brackets in Eqn.~(\ref{eqn:rminusNNLO}) becomes
\begin{eqnarray}
   \left[ 
      \frac{8704}{1215} - \frac{5}{9} \zeta_2 - \frac{16}{45} \zeta_3
      - \frac{83}{162} n_f \right] \, .
\end{eqnarray}
Numerically, for typical values of $\alpha_s$, the NNLO corrections in 
both schemes, $\overline{MS}$ and DIS, are of the order of 30-40 \% of
the NLO contributions. This shows good perturbative stability of the
QCD prediction with respect to higher orders.
We have to note however, that a consistent NNLO
analysis requires NNLO evolution of the corresponding parton
distributions. The necessary three-loop anomalous dimensions are not
completely known yet~\cite{vanNeerven:2000wp,Moch:2002sn}.
 
Thus, we can conclude the pQCD corrections to $R^-$ are small and that
uncertainties due to the perturbative expansion can be reasonably estimated. 

\subsection{Experimental cuts}
\label{sect:expModel}

Because of NuTeV's inability to measure NC events down to zero
inelasticity, NuTeV cannot report ratios of total cross sections.  One
concern, then, is that $R^-$ may not accurately reflect NLO QCD 
corrections to the NuTeV result.  In particular, these cuts either
remove or change the contributions near the kinematic endpoints in the
inelasticity, $y$, which is a reason in general where one might expect
enhanced sensitivity to radiative corrections.

Backgrounds from sources other than muon neutrino interactions can be
reliably subtracted in the NuTeV analysis; however, three major
corrections to $R^-$ remain.  These are: (1) the minimum $\nu$ cut
required in order to observe the hadronic recoil in the detector,
common to both the charged and neutral current, (2) the difference in
the observed visible energy in the charged and neutral current due to
the presence of the final state muon, and (3) charged-current events
at very low $E_\mu$ which fake neutral currents.  To a significant
extent, all of these effects can be modeled by considering the
relation (\ref{eqn:rminus}) with cuts on the inelasticity $y$.

To that end, let us define for any difference of
cross-section the cuts in $y$ as
\begin{eqnarray}
  \label{eqn:ycutcrsdiff}
\left\{\sigma(\nu_{\mu}N\rightarrow X) -
\sigma(\nub_{\mu}N\rightarrow X) \right\} \biggr|_{y_{\rm min}}^{y_{\rm max}}
 &=& 
\int\limits_{y_{\rm min}}^{y_{\rm max}}\, dy\, 
\frac{d\sigma(\nu_{\mu}N\rightarrow X)}{dy} -
\frac{d\sigma(\nub_{\mu}N\rightarrow X)}{dy}
\nonumber \\ 
\end{eqnarray}
with (standard) definition of the inelasticity $y$ being the fraction of
the neutrino's energy lost in the nucleon's rest frame.

A simulation of the effect of the NuTeV broadband flux
translates the experimental minimum cut in visible calorimeter energy
of $20$~GeV into an effective minimum $y$-cut, $y_{\rm min}$, of
$0.24$.  Simulations of the energy deposited by the final state muon
show a reduction of this effective minimum $y$ for charged-current
events by $\delta_y$ of $0.03$.  Finally, a detailed detector and flux
simulation~\cite{Zeller:2001hh} can be used to measure the effective
$y$ at which the final state muons in charged-current events are so
soft that the events are mistaken for neutral-currents.  Denoting this
$y$ at which cross-talk occurs as $1-y_{\rm X}$, $y_{\rm X}$ is
numerically $0.043$.

This motivates the definition of a simple cross-section model for $R^{-}$ as 
 \begin{eqnarray}
{\lefteqn{
R^{-}_{\rm model}(y_{\rm min},\delta y,y_{\rm X}) \equiv}} 
\label{eqn:rminusXtalk}
\\[2ex] &&
\nonumber
\frac{\displaystyle
\left\{\sigma(\nu_{\mu}N\rightarrow\nu_{\mu}X) -
\sigma(\nub_{\mu}N\rightarrow\nub_{\mu}X) \right\} \biggr|_{y_{\rm min}}^{1}
+
\displaystyle
\left\{\sigma(\nu_{\mu}N\rightarrow\mu^-X) -
\sigma(\nub_{\mu}N\rightarrow\mu^+X)  \right\} \biggr|_{1-y_{\rm X}}^{1}}
{\displaystyle
\left\{\sigma(\nu_{\mu}N\rightarrow\mu^-X) -
\sigma(\nub_{\mu}N\rightarrow\mu^+X) \right\} \biggr|_{y_{\rm
min}-\delta y}^{1-y_{\rm X}}}\,
\end{eqnarray}
which accounts for the kinematic cuts discussed.

We can now work out the structure of $R^{-}_{\rm model}$ including 
higher order QCD corrections. 
The result for $R^{-}_{\rm model}$ up to NLO can be written as 
follows (an extension to NNLO is straight forward):
 \begin{eqnarray}
R^{-}_{\rm model}(y_{\rm min},\delta y,y_{\rm X})
 &=& f_{0}(y_{\rm min},\delta y,y_{\rm X})(g_L^2 - g_R^2) 
 + f_{1}(y_{\rm min},\delta y,y_{\rm X})
+ \frac{u^- - d^- + c^- - s^-}{u^- + d^-} 
 \nonumber\\
& &
\times \Biggl( f_u(y_{\rm min},\delta y,y_{\rm X})(g_{Lu}^2 - g_{Ru}^2) 
\nonumber\\
& & 
+ f_d(y_{\rm min},y_{\rm X})(g_{Ld}^2 - g_{Rd}^2) 
+ f_{2}(y_{\rm min},\delta y,y_{\rm X}) 
\nonumber\\
& & 
+ \frac{\alpha_s}{\pi} f_{\alpha_s}(y_{\rm min},\delta y,y_{\rm X})(g_L^2 - g_R^2) 
+ \frac{\alpha_s}{\pi} f_{3}(y_{\rm min},\delta y,y_{\rm X})\Biggr)
\nonumber\\
& &
 + {\cal{O}}\left(\frac{1}{(u^- + d^-)^2}\right) 
 + {\cal{O}}\left(\alpha_s^2\right),
\label{eqn:rXtalkNLOycut}
\end{eqnarray}
where the functions $f_0, \dots ,f_3,f_u,f_d$ and $f_{\alpha_s}$ are
given as 

 \begin{eqnarray}
f_0(y_{\rm min},\delta y,y_{\rm X}) &=& 
{{(y_{\rm min}-1)\*(y_{\rm min}^2-2\*y_{\rm min}-2)}\over
{P_1\*P_2}}
\, ,
\nonumber\\
f_1(y_{\rm min},\delta y,y_{\rm X}) &=& 
- {{y_{\rm X}\*(y_{\rm X}^2-3)}\over
{P_1\*P_2}}
\, ,
\nonumber\\
f_u(y_{\rm min},\delta y,y_{\rm X}) &=& 
-6\*{{(y_{\rm min}-1)\*(y_{\rm min}^2-2\*y_{\rm min}-2)}\over
{P_1\*P_2^2}}
\, ,
\nonumber\\
f_d(y_{\rm min},\delta y,y_{\rm X}) &=& 
-2\*{{(y_{\rm min}-1)\*(y_{\rm min}^2-2\*y_{\rm min}-2)}\over 
{P_1\*P_2^2}}
\nonumber\\
& & 
\times\*(y_{\rm min}^2+y_{\rm
      X}^2+y_{\rm X}+y_{\rm X}\*\delta y+(1+\delta y)^2-y_{\rm min}\*(2+y_{\rm X}+2\delta y))\, ,
\nonumber\\
f_2(y_{\rm min},\delta y,y_{\rm X}) &=& 
-6\*{{(y_{\rm min}-1-\delta y)\*y_{\rm X}\*(-y_{\rm X}+y_{\rm
min}-1-\delta y)}\over
{P_1\*P_2^2}}
\, ,
\nonumber\\
f_{\alpha_s}(y_{\rm min},\delta y,y_{\rm X}) &=& 
-{1\over 9}\*{{(y_{\rm min}-1)\*(y_{\rm min}^2-2\*y_{\rm
      min}-2)}}\over 
{P_1\*P_2^2}
\nonumber\\
& & \times\*\Biggl( y_{\rm min}^2-y_{\rm min}\*y_{\rm X}
      -2y_{\rm min}\*\delta y-14\*y_{\rm
    min} 
\nonumber\\
& &
+(13+\delta y)y_{\rm X}+16+y_{\rm X}^2+14\delta y+\delta y^2\Biggr) 
\, ,
\nonumber\\
f_{3}(y_{\rm min},\delta y,y_{\rm X}) &=& 
-{2\over 3}\*{{(y_{\rm min}-1-\delta y)\*y_{\rm X}}\over 
{P_1\*P_2^2}}
\nonumber\\
& &
\times\*(-3\*y_{\rm X}+2\*y_{\rm min}\*y_{\rm X}-7+y_{\rm
  min}-\delta y-2y_{\rm X}\delta y)
\, ,
\label{eqn:ycutg}
\end{eqnarray}
\noindent
where
\begin{eqnarray}
P_1 &=& \denone\nonumber\\
P_2 &=& \dentwo
\, .
\end{eqnarray}
\noindent
In the limit $y_{\rm min},\delta y,y_{\rm X} \to 0$ the functions
$f_0, \dots ,f_3,f_u,f_d$ and $f_{\alpha_s}$ simplify to 
\begin{eqnarray}
&
f_0(y_{\rm min},\delta y,y_{\rm X}) = 1,\qquad 
f_1(y_{\rm min},\delta y,y_{\rm X}) = 0,
\nonumber\\
&\displaystyle
f_2(y_{\rm min},\delta y,y_{\rm X}) = 0,\qquad
f_3(y_{\rm min},\delta y,y_{\rm X}) = 0,  
\nonumber\\
&\displaystyle
f_u(y_{\rm min},\delta y,y_{\rm X}) = 3,\qquad
f_d(y_{\rm min},\delta y,y_{\rm X}) = 1,
\nonumber\\
&\displaystyle
f_{\alpha_s}(y_{\rm min},\delta y,y_{\rm X}) =  {8\over 9},  
\label{eqn:ycutspecfunc}
\end{eqnarray}
in agreement with Eqn.~(\ref{eqn:rminusNLO}).  Putting in the
numerical values from the experimental $y$ cuts, we find
\begin{eqnarray}
&
f_0(y_{\rm min},\delta y,y_{\rm X}) = 1.053,\qquad 
f_1(y_{\rm min},\delta y,y_{\rm X}) = 0.074,
\nonumber\\
&\displaystyle
f_2(y_{\rm min},\delta y,y_{\rm X}) = 0.042,\qquad
f_3(y_{\rm min},\delta y,y_{\rm X}) = 0.038,  
\nonumber\\
&\displaystyle
f_u(y_{\rm min},\delta y,y_{\rm X}) = 2.700,\qquad
f_d(y_{\rm min},\delta y,y_{\rm X}) = 0.594,
\nonumber\\
&\displaystyle
f_{\alpha_s}(y_{\rm min},\delta y,y_{\rm X}) =  0.683,  
\label{eqn:ycutspecfuncexact}
\end{eqnarray}
\noindent
Numerically, then, the NLO QCD correction to $R^{-}_{\rm model}$ is
$-0.00033$.  For comparison, the value for $dR^{-}_{\rm model}/d\stw$ is $-1.0075$.
In summary, in this model the experimental cuts in $y$ do not
result in large increases in the ${\cal O}(\alpha_s)$ QCD corrections.

\subsection{Parton Distribution Functions}
\label{sect:PDFs}

Let us next discuss the parton distributions. The very fact, that
corrections to $R^-$ in the QCD improved parton model are proportional 
to isotriplet or $C$-odd components of the nucleon target has led
to questions about our knowledge of parton distributions for the 
various quark flavors.  There are two major issues: the isotriplet
component introduced by the small neutron excess of the NuTeV
target, and possible momentum asymmetries between the strange and
anti-strange seas.

The NuTeV analysis corrects for the significant asymmetry of $d$ and
$u$ quarks that arises because the NuTeV target, which is primarily
composed of iron, has a $\delta N\equiv (A-2Z)/A = +0.0574\pm0.0002$
fractional excess of neutrons over protons.  As can be observed from
the above uncertainty, the neutron excess is very well known due to a
detailed material survey of the NuTeV target, including a chemical
assay of the NuTeV
steel~\cite{Zeller:2002du,McFarland:2002sk,Zeller:2002he}.  The
largest uncertainty in this isotriplet correction, in fact, comes from
the difference between $u$ and $d$ PDFs.  From the uncertainty on the
NMC $F_2^d/F_2^p$~\cite{Arneodo:1996kd}, NuTeV estimates this
uncertainty to be $0.0003$ in their extraction of $\stw$.  (Note that
this correction assumes isospin symmetry, i.e.,
$\uubar_p(x)=\ddbar_n(x)$, $\ddbar_p(x)=\uubar_n(x)$, violations of
which are discussed below.)

Kulagin~\cite{Kulagin:2003wz} has recently noted that NuTeV's
measurement of the neutron excess correction in $\stw$, $0.0080$, does
not agree with the correction of the effect in $R^-$, and suggests
that the difference could be taken as a (substantial) correction to
the NuTeV measurement.  However, the differences can be partly understood
in terms of the experimental cuts discussed in
Section~\ref{sect:expModel} which reduce the neutron excess correction
by $20\%$, and partly understood as the differences between the NuTeV
1C fit and $R^-$.  Therefore, the suggestion by Kulagin that this
difference is a correction to NuTeV's $\stw$ is incorrect.

Global analyses of unpolarized parton distributions 
usually assume no strange asymmetry, i.e. imposing as a constraint 
$s(x) = \overline{s}(x)$.
Moreover, a fit ansatz like \cite{Pumplin:2002vw}
\begin{eqnarray}
  \label{eqn:cteqansatz}
  s(x) = \overline{s}(x) = \kappa\, {\overline{u}(x) +
  \overline{d}(x) \over 2}\, ,
\end{eqnarray}
ties the strangeness distribution to the relatively well known
$\overline{u}(x)$ and $\overline{d}(x)$ distributions, thereby
underestimating the true uncertainty~\cite{olness:dis03}.
There are several recent dedicated analyses of inclusive lepton-nucleon DIS
(including neutrino data)
\cite{Barone:1999yv,Botje:1999dj,Alekhin:2000ch}, which have used
various methods to account for the uncertainties and correlated errors
in parton distributions \cite{Botje:1999dj}, \cite{Giele:2001mr}.

One of these
analyses~\cite{Barone:1999yv} of the CDHS neutrino data and
charged-lepton data reported some improvement in their fits if they
allow for an asymmetry in the strange sea at high $x$.  However, this
large asymmetry at high $x$ is directly excluded by the
non-observation of high $x$ dimuon events at NuTeV and
CCFR\cite{Goncharov:2001qe}.  An update of this analysis using CCFR
and CDHS charged-current neutrino data
no longer finds the large significant asymmetry between the strange
and anti-strange quark distributions~\cite{portheault:dis03}.

As illustrated above, the NuTeV dimuon data \cite{Goncharov:2001qe}
can help in understanding the quark flavor content of the nucleon as
it provides separated measurements of $s(x)$ and ${\overline{s}}(x)$
initial state contributions to charm production.

NuTeV has measured possible differences between $s(x)$ and
${\overline{s}}(x)$ from the CCFR and NuTeV data on $\txnunub N\to
\mu^+\mu^- X$ within the NuTeV enhanced LO cross-section model used in
the $\stw$ analysis.  Denote the momentum integrals of the strange and anti-strange seas as $S$, $\Sbar$, i.e., $S=\int xs(x)dx$ and $\Sbar=\int x\sbar(x)dx$.
Within this model, the dimuon data implies a
{\em negative} asymmetry\cite{Zeller:2002du},
\begin{equation}
 S-\Sbar = -0.0027 \pm 0.0013,
\end{equation}
\noindent
or an asymmetry of $11\pm6$\% of $(S+\Sbar)$.
Therefore, dropping the assumption of strange-antistrange symmetry results in
an {\em increase} in the NuTeV value of $\stw$,
\begin{equation}
 \Delta\stw = +0.0020 \pm 0.0009.
\end{equation}
\noindent
The initial NuTeV measurement, which assumes $s(x)=\sbar(x)$, becomes 
\begin{equation}
\stwos=0.2297\pm0.0019.
\end{equation}
\noindent 
A preliminary analysis of the strange and anti-strange asymmetry in an
NLO cross-section model also finds the momentum carried by the seas to
be consistent within uncertainties.  It is worth noting that these
fits have been criticized in the literature~\cite{Davidson:2001ji}
because of their assumed functional form in $x$, and indeed, NuTeV is
planning to reanalyze this data with functional forms more suggestive
of strange and anti-strange asymmetries suggested by non-perturbative
calculations~\cite{Signal:1987gz,Burkardt:1991di,Brodsky:1996hc,Melnitchouk:1999mv}.

\begin{figure}[tbp]
\begin{center}
\epsfxsize=0.8\textwidth\epsfbox{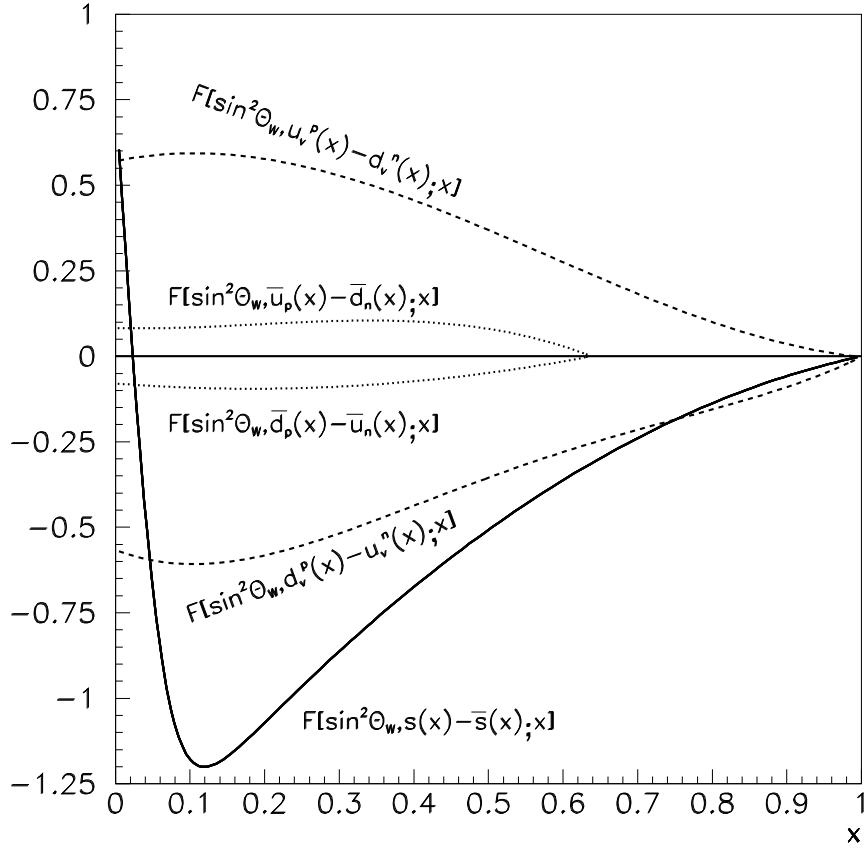}
\end{center}
\caption{\it
  The functionals describing the shift in the NuTeV $\stw$ caused by
  not correcting the NuTeV analysis for isospin violating $u$ and $d$
  valence and sea distributions or for $\sav\neq\savbar$.  The shift in
  $\stw$ is determined by convolving the asymmetric momentum distribution with
  the plotted functional.}
\label{fig:stw}
\end{figure}

The impact of uncertainties in the strange-antistrange and in the up-down
quark asymmetries of the target on the NuTeV measurement of $\sin^2
\theta_W$ has been quantified in terms of an error
functional~\cite{Zeller:2002du},  $F[\stw,\delta; x]$ such that 
\begin{equation}
\Delta\stw = \int_0^1 F[\stw,\delta;x]\,\delta(x)\:dx,
\end{equation}
for any symmetry violation $\delta(x)$ in PDFs.  All of the details of
the NuTeV analysis are included in the numerical evaluation of the
functionals shown in Figure~\ref{fig:stw}.  For this analysis, it can
be seen that the level of isospin violation required to shift the
$\stw$ measured by NuTeV to its standard model expectation would be,
e.g., $\int x(d_p(x)-u_n(x))dx\sim0.01$ (about 5\% of $\int
x(d_p(x)+u_n(x))dx$), and that the level of asymmetry in the strange
sea required would be $S-\Sbar\sim +0.007$ (about $30\%$ of
$S+\Sbar$).

\subsection{Charged current charm production}

Thus far, the discussion of QCD corrections to $R^-$ as in
Eqns.~(\ref{eqn:rminusNLO}), (\ref{eqn:rminusNNLO}) has assumed quarks
to be massless.  However, an important contribution to the charged
current reactions is the deep-inelastic charm production in $\nu s \to
\mu^- c X$. Clearly, the mass $m_{\rm c}$ of the charm quark should to
be taken into account.

For the center-of-mass energies accessible to NuTeV, a scheme of
three light flavors in the nucleon is expected to give 
an appropriate approximation. Thus, charm quarks are entirely 
generated perturbatively. 
The leading order QCD improved parton model then requires the
$x$-dependence to change slightly. 
The relevant strange distribution becomes $s(\xi,\mu_{\rm f}^2)$, 
where $\xi=x(1+m_{\rm c}^2/Q^2)$ as the scaling variable correctly 
accounts for the single-charm threshold condition.
This amounts to the so-called ``slow rescaling'' \cite{Gottschalk:1981rv}. 
This, for instance, has been implemented in the NuTeV analysis 
of the dimuon production cross-section \cite{Goncharov:2001qe}.

Also the higher order QCD corrections to $\nu s \to \mu^- c X$ are 
known since long. At order $\alpha_s$, the complete
corrections to deep-inelastic charged current scattering 
have been calculated~\cite{Gottschalk:1981rv,Gluck:1996ve}. From 
these results, it is straight forward to obtain the complete
$m_{\rm c}$-dependence of $R^-$ at NLO. However, for the purpose of this 
letter we restrict ourselves to a discussion of certain qualitative 
features.
First of all, for an observable like $R^-$, consisting of a particular 
combination of total cross-sections, some dependence on charm mass cancels. 
Most prominently, there will be no large logarithms $\ln(Q^2/m_{\rm c}^2)$ 
in the NLO corrections to $R^-$ due to the combination of 
coefficient functions in Eqn.~(\ref{eqn:coefffct}). 
In comparison with Eqn.~(\ref{eqn:rminusNLO}), 
the treatment of charm as a massive quark will at most 
introduce additional terms of order $m_{\rm c}^2/Q^2$ in the coefficients 
proportional to $\alpha_s$. 

At order $\alpha_s^2$, only those corrections 
for deep-inelastic charged current scattering are known, which are 
logarithmically enhanced by ${\cal{O}}(\alpha_s^2 \ln^n(Q^2/m_{\rm c}^2))$ 
contributions \cite{Buza:1997mg}. For $R^-$ this implies, that 
there can be at most single logarithms $\ln(Q^2/m_{\rm c}^2)$ at NNLO 
due to the absence of logarithmic enhanced terms at NLO.

Thus, we conclude that for $R^-$, any dependence on the charm mass 
should be weak. The same conclusion also holds, if we model the effect 
of experimental cuts like in $R^{-}_{\rm model}$ of 
Eqn.~(\ref{eqn:rminusXtalk}). There again, the $m_{\rm c}$-dependence
enters only through additional terms of order $m_{\rm c}^2/Q^2$.

However, as already mentioned in Section~\ref{sect:notR-}, the charm
mass dependence does play a role in choosing how the experimental
$\Rmeasnu$ and $\Rmeasnub$ are combined to extract $\stw$.  Although
the result is robust over changes in this prescription, this
dependence is still worth noting and may bear further investigation.
Also, the extraction of the strange sea from the CCFR/NuTeV dimuon
data, which is used as input to the $\stw$ measurement has significant
NLO corrections.  Again, the (quark-antiquark symmetric) strange sea
doesn't enter directly into $R^-$ but does enter into the experimental
combinations used in the NuTeV fits.

For a full quantitative analysis of the various experimental
cuts in phase space a detailed Monte Carlo study has to be performed.
This has to account also for fragmentation, which has been entirely 
neglected in this discussion. In particular, the fragmentation has a
non-trivial effect on the visible final state energy due to the
presence of the hard neutrino from the charm decay. 
A Monte Carlo program, which calculates the fully differential 
cross-sections has been provided in \cite{Kretzer:2001tc}.

\subsection{Isospin breaking}
\label{sect:iso}

Having concluded the investigation of perturbative QCD corrections to
$R^{-}$ let us now turn to investigate potential nuclear effects and 
isospin violation. 

As argued above, knowledge of the neutron excess allows for a
reasonably accurate correction for the isovector part of the
cross-section; however, this correction is only valid with the
assumption of isospin symmetry, i.e., $\uubar_p(x)=\ddbar_n(x)$,
$\ddbar_p(x)=\uubar_n(x)$. This assumption, if significantly
incorrect, could produce a sizable effect in the NuTeV extraction of
$\stw$~\cite{Sather:1992je,Rodionov:1994cg,Cao:2000dj,Davidson:2001ji,Zeller:2002du,Kulagin:2003wz}.

Let us briefly discuss the main proposed non-perturbative models to
generate isospin violation in the
nucleon~\cite{Sather:1992je,Rodionov:1994cg,Cao:2000dj}.  The earliest
estimate in the literature, a bag model calculation by
Sather~\cite{Sather:1992je}, predicts large valence asymmetries of
opposite sign in $u_p-d_n$ and $d_p-u_n$ at all $x$, but neglects a
number of effects.  Most notably, the effective mass of the remnant
diquark is assumed to have a $\delta$-function distribution.
Recently, Londgergan and Thomas, revisiting their earlier
calculation~\cite{Rodionov:1994cg} with the fixed diquark mass in the
bag model but including a number of effects neglected by Sather,
including nucleon size and mass, have argued that their calculation
observes the same effect as Sather~\cite{Londergan:2003ij} and that
this effect is largely independent of PDFs~\cite{Londergan:2003pq}, at
least when the $x$ dependence of the NuTeV
acceptance~\cite{Zeller:2002du} is neglected.  However, when this same
calculation is done with a smeared distribution of diquark
masses~\cite{Rodionov:1994cg}, the dominant isospin violating effect
of the minority quark distribution, $d_p-u_n$, is reduced at high $x$
and the negative asymmetry at low $x$ is found to carry more momentum.
Thus including the effect of diquark smearing, the high $x$ and low
$x$ contributions largely cancel, leaving a negligible ($0.0001$)
shift in the NuTeV $\stw$~\cite{Zeller:2002du}.  The effect is also
evaluated in the meson cloud model~\cite{Cao:2000dj}, and there the
asymmetries are much smaller at all $x$, again resulting only in a
small shift in $\stw$.

Models aside, the NuTeV data itself cannot provide a significant
independent constraint on this form of isospin violation. However, if
parton distributions extracted from neutrino data (on heavy targets)
are used to separate sea and valence quark distributions which affect
observables at hadron colliders~\cite{Bodek:1999bb}, effects of
isospin breaking could be seen.  Therefore, global analyses of parton
distributions including the possibility of isospin violation may be
able to constrain this possibility further experimentally.

\subsection{Nuclear effects}

Any nuclear effect which can be
absorbed into process-independent PDFs will not affect the NuTeV
result.  However, several authors have recently explored the
possibility that neutrino neutral and charged current reactions may
see different nuclear effects and therefore influence the NuTeV
result.

\begin{figure}[tbp]
\epsfxsize=\textwidth\epsfbox{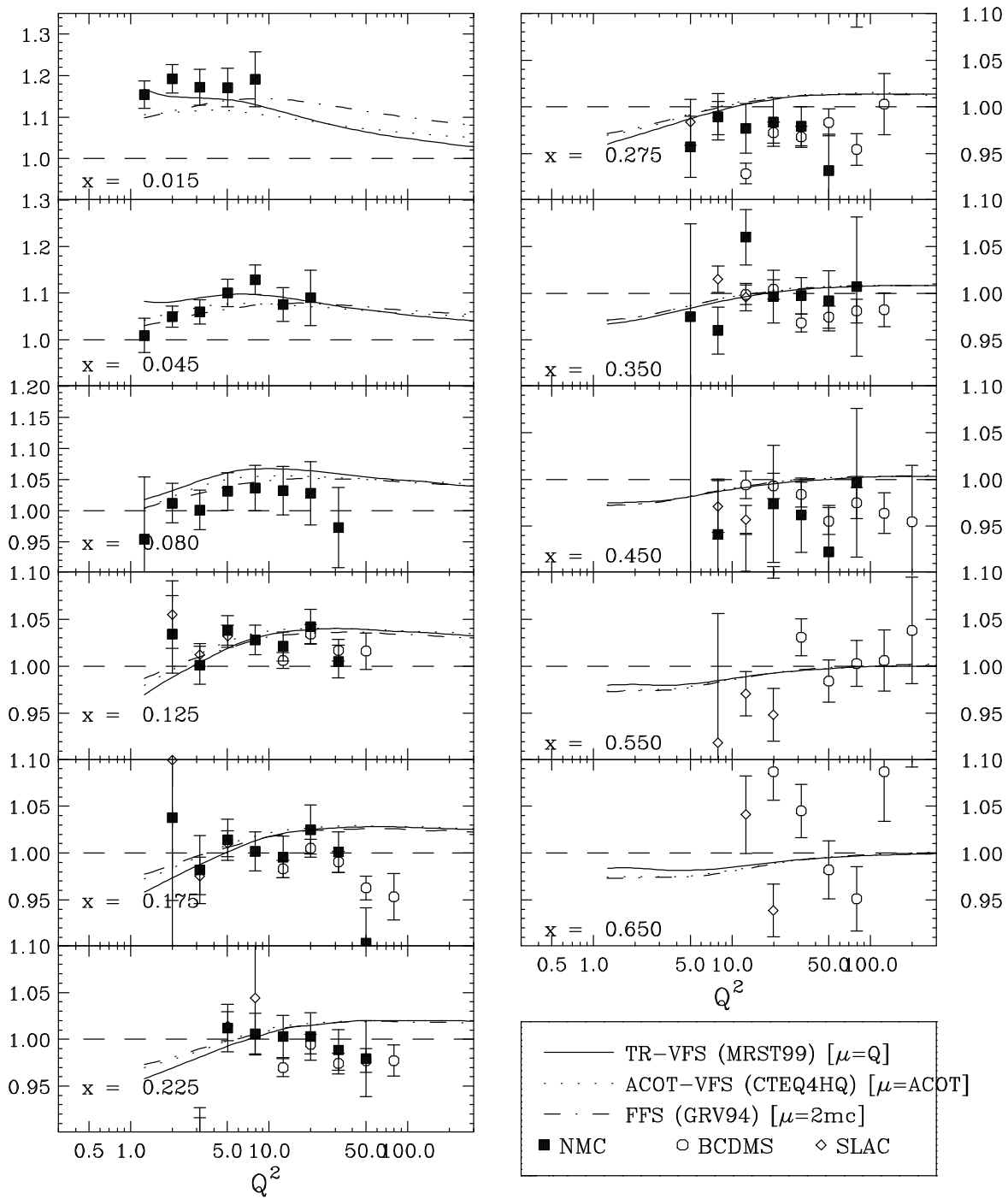}
\caption{\it
  The ratio of $F_2^{\nu CC}$ with a model-independent
  prediction from $F_2^{\ell^\pm}$ on iron.  Primarily because
  of quark-mass effects these ratios are not expected to be one,
  and different theoretical calculations for this ratio are shown.
  Good agreement of the ratio with expectations limits anomalous
  nuclear effects in the $\nu$ charged-current interaction.}
\label{fig:ccfr-pmi}
\end{figure}

Thomas and Miller~\cite{Miller:2002xh} have offered a Vector
Meson Dominance (VMD) model of low $x$ shadowing in which such an
effect might arise.  The NuTeV analysis, which
uses $\nu$ and $\nubar$ data at $<Q^2>$ of $25$ and $16$~GeV$^2$,
respectively, is far away from the VMD regime, and the effect of this
VMD model is significantly smaller than stated in
this analysis.  The most serious flaw in the hypothesis
that this accounts for the NuTeV result, however, is that it is not
internally consistent with the NuTeV data. Shadowing, a low $x$
phenomenon, largely affects the sea quark distributions which are
common between $\nu$ and $\nubar$ cross-sections, and therefore cancel
in $R^-$.  However, the effects in $\Rnu$ and $\Rnub$ individually are
much larger than in $R^-$ and this model {\em increases} the
prediction for NuTeV's $\Rnu$ and $\Rnub$ by $0.6\%$ and $1.2\%$,
respectively.  NuTeV's $\Rnu$ and $\Rnub$ are both below predictions
and the significant discrepancy is in the $\nu$ mode, not the $\nubar$
``control'' sample, both in serious contradiction with the prediction
of the VMD model~\cite{Zeller:2002et}.

Kulagin~\cite{Kulagin:2003wz} has recently investigated possibilities
for process-dependent nuclear effects disrupting NuTeV due to Fermi
motion, nuclear binding corrections and shadowing, and found the
effects to be small. Schmidt and collaborators~\cite{Kovalenko:2002xe}
have suggested that there may be little or no EMC effect in the
neutrino charged current but the expected EMC effect suppression at
high $x$ in the neutral current.  If true, this could have the right
behavior and perhaps magnitude to explain the NuTeV data because of
the effect at high $x$.  Unfortunately, this mechanism would cause
large differences between $F_2^\nu$ and $F_2^{\ell}$ on heavy targets
at high $x$ which are excluded by the CCFR charged current
cross-section measurements~\cite{Yang:2000ju} shown in
Figure~\ref{fig:ccfr-pmi}.

Kumano~\cite{Kumano:2002ra} has fit experimental
DIS and Drell-Yan data on nuclear targets to investigate the
possibility that nuclear effects are flavor dependent, rather than
process dependent.  Such an effect could impact NuTeV because the
constraints on $d/u$ of the nucleon come from light targets.  Kumano
performs these fits in the context of ``nuclear PDFs'', and found very
small flavor-dependent effects, except at very high $x$ and low $Q^2$,
a region removed by the visible energy requirement ($E_{\rm
calorimeter}>20$~GeV) of the NuTeV analysis.  The effect on the NuTeV
analysis which assumes flavor-independent nuclear effects is therefore
negligible.

\section{Conclusions}

Motivated by the NuTeV measurement of $\stw$, which deviates from 
the standard model prediction by approximately 3~$\sigma$, 
we have studied potential sources of theoretical uncertainties.
In particular, we have investigated QCD effects at higher orders 
in perturbation theory including the dependence on the charm mass.
We have discussed the impact of parton distributions and potential 
nuclear effects. To assess the effect of experimental cuts on the 
measured $R^-$, we have developed a simple model $R^{-}_{\rm model}$ 
which accounts for $y$-cuts up to NLO perturbative QCD.

Based on these investigations, we conclude that higher QCD corrections
are under control and small. The uncertainties on parton distributions
have been discussed and will also be addressed in future global
analyses.  An asymmetric strange sea seems unlikely to be an
explanation of the present discrepancy.  Large isospin violation in
parton distribution functions is a possible explanation, but the
violation would have to be larger than naive estimates would suggest.

\section*{Acknowledgements}

We thank Stefano Forte and Geralyn Zeller for useful input to this
work.  We also thank Bogdan Dobrescu, Keith Ellis, Paolo Gambino,
Sergei Kulgain, Dave Mason, Fred Olness, and Benjamin Porthault for
useful discussions that have shaped the conclusions of this work.  One
of us (KSM) acknowledges support from the United States Department of
Energy, the United States National Science Foundation and the Research
Corportation of Tucson, Arizona, USA.

\section*{Appendix}

Here we give some formulae relevant for the calculation of QCD
corrections to $R^-$.
Neutral/charged current cross-sections are defined as \cite{Hagiwara:2002fs}
\begin{eqnarray}
\label{eq:NCCCcrssect}
{d^2 \sigma^{\rm NC/CC} \over dx dy} &=& {G_F^2 \over \pi}\, {s \over 1+Q^2/M_Z^2}\, k^{\rm NC/CC}\, ,
\end{eqnarray}
with 
\begin{eqnarray}
Y^+= (1+(1-y)^2)/2\, \qquad\qquad Y^- = (1-(1-y)^2)/2\, . 
\end{eqnarray}
The $k^{\rm NC/CC}$ in Eqn.~(\ref{eq:NCCCcrssect}) are expressed in 
terms of the structure functions $F_2,F_3, F_L$
\begin{eqnarray}
k^{{\rm NC}\nu}&=&Y^+\, F_2^{\rm NC}+Y^-\, xF_3^{\rm NC}-y^2/2 F_L^{\rm NC},
\nonumber\\
k^{{\rm NC}\overline{\nu}}&=&Y^+\, F_2^{\rm NC}-Y^-\, xF_3^{\rm NC}-y^2/2 F_L^{\rm NC},
\nonumber\\
k^{{\rm CC}\nu}&=&Y^+\, F_2^{\rm CC}+Y^-\, xF_3^{\rm CC}-y^2/2 F_L^{\rm CC},
\nonumber\\
k^{{\rm CC}\overline{\nu}}&=&Y^+\, F_2^{\rm \overline{C}C}-Y^-\,xF_3^{\rm \overline{C}C}
                           -y^2/2 F_L^{\rm \overline{C}C}.
\end{eqnarray}

For the Paschos-Wolfenstein $R^-$, the discussion can be restricted to the non-singlet
structure functions. In the QCD improved parton model (assuming
massless quarks), these are given as convolutions of parton distributions and 
the coefficient functions of the hard scattering process (see for
instance \cite{Furmanski:1982cw,Zijlstra:1992qd} or {\it Handbook of
  perturbative QCD} \cite{Brock:1995sz}). 
\begin{eqnarray}
F_{i}^{\rm NC}(x) &=&x \int_x^1 \frac{dz}{z}\, \left\{
(u_L^2+u_R^2) (u(z)+{\overline{u}}(z)+c(z)+{\overline{c}}(z)) \right.
\nonumber \\
& & \left.
+
(d_L^2+d_R^2) (d(z)+{\overline{d}(z)}+s(z)+{\overline{s}}(z))\right\}
C_{i,\rm{q}}(x/z)\, ,
\nonumber\\
xF_3^{\rm NC}(x)&=&x \int_x^1 \frac{dz}{z}\, \left\{(u_L^2-u_R^2)
  (u(z)-{\overline{u}}(z)+c(z)-{\overline{c}}(z)) \right.
\nonumber \\
& & \left.
+
(d_L^2-d_R^2) (d(z)-{\overline{d}}(z)+s(z)-{\overline{s}}(z))\right\}
C_{3,\rm{q}}(x/z)\, ,
\nonumber\\
F_{i}^{\rm CC}(x) &=&x \int_x^1 \frac{dz}{z}\,
\left\{{\overline{u}}(z)+d(z)+s(z)+{\overline{c}}(z)\right\}
C_{i,\rm{q}}(x/z)\, ,
\nonumber\\
xF_3^{\rm CC}(x) &=&x \int_x^1 \frac{dz}{z}\,
\left\{-{\overline{u}}(z)+d(z)+s(z)-{\overline{c}}(z)\right\}
C_{3,\rm{q}}(x/z)\, ,
\nonumber\\
F_{i}^{\rm \overline{C}C}(x) &=&x \int_x^1 \frac{dz}{z}\,
\left\{u(z)+{\overline{d}}(z)+{\overline{s}}(z)+c(z)\right\}
C_{i,\rm{q}}(x/z)\, ,
\nonumber\\
xF_3^{\rm \overline{C}C}(x) &=&x \int_x^1 \frac{dz}{z}\,
\left\{u(z)-{\overline{d}}(z)-{\overline{s}}(z)+c(z)\right\}
C_{3,\rm{q}}(x/z)\, ,
\label{pqcdstructfct}
\end{eqnarray}
where $i=2,L$ and the non-singlet coefficient functions have an expansion in powers of
$\alpha_s$,
\begin{eqnarray}
  C_{2,\rm{q}}(x) &=& \delta(1-x) + {\alpha_s \over 4\pi} c_{2,\rm q}^{(1)}(x) 
+ {\alpha_s^2 \over (4\pi)^2} c_{2,\rm q}^{(2)}(x)  + \dots 
\nonumber \\
  C_{3,\rm{q}}(x) &=& \delta(1-x) + {\alpha_s \over 4\pi} c_{3,\rm q}^{(1)}(x) 
+ {\alpha_s^2 \over (4\pi)^2} c_{3,\rm q}^{(2)}(x)  + \dots 
\nonumber \\
  C_{L,\rm{q}}(x) &=& {\alpha_s \over 4\pi} c_{L,\rm q}^{(1)}(x) 
+ {\alpha_s^2 \over (4\pi)^2} c_{L,\rm q}^{(2)}(x)  + \dots 
\label{eq:coefffct}
\end{eqnarray}
If restricted to leading order, these equations reproduce the naive
quark parton model. 

In Eqns.~(\ref{pqcdstructfct}), (\ref{eq:coefffct}) all scale
dependence has been suppressed. The structure functions, being 
observables, depend on the physical scale $Q$, 
whereas the PDFs and the coefficient functions depend on the 
renormalization scale $\mu_{\rm r}$ and the factorization scale
$\mu_{\rm f}$, the usual choice being $\mu=\mu_{\rm r}=\mu_{\rm f}$.
Putting additionally $\mu=Q$ also cancels all scale dependent 
logarithms in the coefficient functions at $l$-loops, i.e. all terms 
$\ln^i(Q^2/\mu^2), i\le l$. 

In the Paschos-Wolfenstein relation $R^-$ only total
cross-sections enter. Here, the expressions in Eqn.~(\ref{pqcdstructfct}) 
simplify considerably. After the $x$-integration to obtain total cross-sections, 
all convolutions become simple products of (Mellin)-moments. 
Thus, 
\begin{eqnarray}
\int_0^1 dx\, F_{i}^{\rm NC} &=& \left\{
(u_L^2+u_R^2) (u+{\overline{u}}+c+{\overline{c}}) \right.
\nonumber \\
& & \left.
+
(d_L^2+d_R^2) (d+{\overline{d}}+s+{\overline{s}})\right\}
C_{i,\rm{q}}\, ,
\\
\int_0^1 dx\, xF_3^{\rm NC}(x)&=&\left\{(u_L^2-u_R^2)
  (u-{\overline{u}}+c-{\overline{c}}) \right.
\nonumber \\
& & \left.
+
(d_L^2-d_R^2) (d-{\overline{d}}+s-{\overline{s}})\right\}
C_{3,\rm{q}}\, ,
\\
\int_0^1 dx F_{i}^{\rm CC}(x) &=& \left\{{\overline{u}}+d+s+{\overline{c}}\right\}
C_{i,\rm{q}}\, ,
\\
\int_0^1 dx\, xF_3^{\rm CC}(x) &=& \left\{-{\overline{u}}+d+s-{\overline{c}}\right\}
C_{3,\rm{q}}\, ,
\label{N2pqcdstructfct}
\end{eqnarray}
where $i=2,L$ and the quantities on the right hand sides 
are the second Mellin moments of PDFs and coefficient functions.
Beyond leading order, the factorization of structure functions into 
PDFs and coefficient functions is arbitrary and introduces  scheme
dependence. The most common schemes are $\overline{MS}$, for which most
higher order cross-sections have been calculated, and the DIS-scheme,
which is physically motivated by demanding 
\begin{eqnarray}
  C_{2,\rm{q}}(x) &=& \delta(1-x)
\end{eqnarray}
to all orders in perturbative QCD. Of course, the PDFs and the 
coefficient functions change accordingly and the implications 
for the Paschos-Wolfenstein relation have been discussed in Section~4.

\end{document}